\documentclass[a4paper,11pt]{article}
\usepackage{pos, wrapfig}

\title{Pion-pole contribution to HLbL from twisted mass lattice QCD at the physical point}
\ShortTitle{Pion-pole contribution to HLbL from TM lattice QCD}

\author*[a]{S.~Burri}
\author[b,c]{C.~Alexandrou}
\author[c]{S.~Bacchio}
\author[d]{G.~Bergner}
\author[c]{J.~Finkenrath}
\author[a]{A.~Gasbarro}
\author[b,c]{K.~Hadjiyiannakou}
\author[e]{K.~Jansen}
\author[f]{B.~Kostrzewa}
\author[c]{G.~Koutsou}
\author[g]{K.~Ottnad}
\author[h,i]{M.~Petschlies}
\author[c]{F.~Pittler}
\author[h,i]{F.~Steffens}
\author[h,i]{C.~Urbach}
\author[a]{U.~Wenger}

\affiliation[a]{Albert Einstein Center for Fundamental Physics,
  Institute for Theoretical Physics, University of Bern,
  Sidlerstrasse 5,
  CH--3012 Bern, Switzerland}

\affiliation[b]{Department of Physics, University of Cyprus, 20537~Nicosia, Cyprus}

\affiliation[c]{Computation-based Science and Technology Research Center, The Cyprus Institute, 20~Konstantinou Kavafi Street, 2121~Nicosia, Cyprus}

\affiliation[d]{University of Jena, Institute for Theoretical Physics, Max-Wien-Platz 1, D-07743~Jena, Germany}

\affiliation[e]{Deutsches Elektronen-Synchrotron DESY, Platanenallee
  6, 15738 Zeuthen, Germany}

\affiliation[f]{High Performance Computing and Analytics Lab,
  Rheinische Friedrich-Wilhelms-Universit{\"a}t Bonn,
  Friedrich-Hirzebruch-Allee~8,
  D-53115 Bonn, Germany}

\affiliation[g]{PRISMA$^+$ Cluster of Excellence and Institut f{\"u}r Kernphysik,
Johannes Gutenberg-Universit{\"a}t Mainz, Johann-Joachim-Becher-Weg
45, D-55128 Mainz, Germany}

\affiliation[h]{HISKP (Theory), Rheinische
  Friedrich-Wilhelms-Universit{\"a}t Bonn,
  Nussallee~14-16,
  D-53115~Bonn, Germany}

\affiliation[i]{Bethe Center for Theoretical Physics, Rheinische
  Friedrich-Wilhelms-Universit{\"a}t Bonn, Wegelerstraße 10, D-53115 Bonn, Germany}

\emailAdd{burri@itp.unibe.ch}

\newcommand{\FPtogg}{{\cal F}_{P\rightarrow \gamma^*\gamma^*}}
\newcommand{\Fpitogg}{{\cal F}_{\pi \rightarrow \gamma^*\gamma^*}}
\newcommand{\fig}{Figure }
\newcommand{\eq}{Eq.~}
\newcommand{\eqs}{Eqs.~}

\abstract{ We report on our computation of the pion transition form
  factor $\Fpitogg$ from twisted mass lattice QCD in order to determine the
  numerically dominant light pseudoscalar pole contribution in the
  hadronic light-by-light scattering contribution to the anomalous
  magnetic moment of the muon $a_\mu =(g-2)_\mu$. The pion transition form
  factor is computed directly at the physical point. We present first
  results for our estimate of the pion-pole contribution with
  kinematic setup for the pion at rest.  }

\FullConference{%
 The 38th International Symposium on Lattice Field Theory, LATTICE2021
  26th-30th July, 2021
  Zoom/Gather@Massachusetts Institute of Technology
}

\begin{document}
\maketitle

\section{Introduction}
In this project we aim to compute the pseudoscalar transition form
factors $\FPtogg$ from twisted mass lattice QCD for the three
pseudoscalar states $P=\pi^0, \eta$ and $\eta'$ in order to determine
the corresponding pseudoscalar pole contributions in 
the hadronic light-by-light (HLbL) scattering contribution to the anomalous
magnetic moment of the muon $a_\mu = (g-2)_\mu$. 
Our computation is done on two
ensembles with the pion mass at its physical value. For our
calculations we are using twisted-mass clover-improved lattice QCD at
maximal twist, so that we have automatic ${\cal O}(a)$-improvement in
place. The generation of the two ensembles was done in the context of
the Extended Twisted Mass Collaboration (ETMC) where the $N_f=2+1+1$ simulations
include the two mass-degenerate light $u$- and $d$-quark flavours at
their physical quark-mass values and
the heavier $s$- and $c$-quark flavours at quark masses close to their
physical values. At the moment, the
analysis is done on two physical point ensembles at two different
lattice spacings as decribed in Table \ref{tab:ensembles}.
For further details on the simulations we refer to
Refs.~\cite{ExtendedTwistedMass:2021gbo,ExtendedTwistedMass:2021qui}.
%
\begin{table}[h]
  \centering
  \begin{tabular}{|l | c | c | c | c | c |}
    \hline
    ensemble & $L^3 \cdot T / a^4$ & $m_\pi$ [MeV] & $a$ [fm] & $L$ [fm] & $m_\pi \cdot L$\\ \hline
    cB072.64 & $64^3 \cdot 128$ & 136.8(6) & 0.082 & 5.22 & 3.6\\
    cC060.80 & $80^3 \cdot 160$ & 134.2(5) & 0.069 & 5.55 & 3.8\\
    \hline
  \end{tabular}
  \caption{Description of ensembles used for the analysis presented in these proceedings.
    \label{tab:ensembles}
  }
\end{table}
%
 
\begin{figure}[b]
  \centering
\includegraphics[height=4cm ]{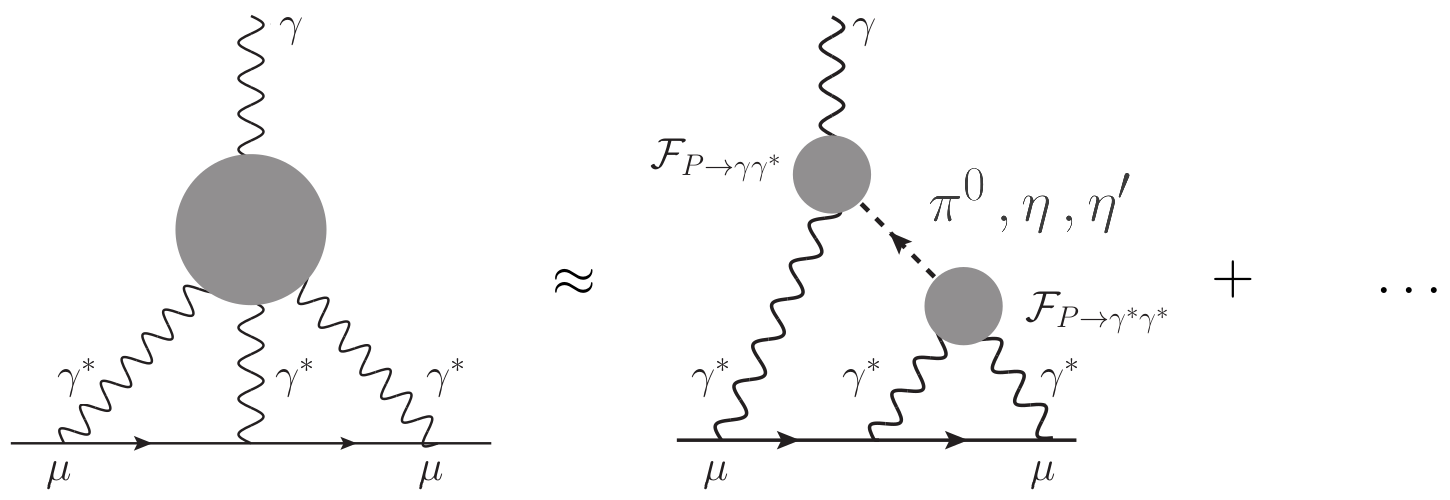} 
\caption{Pseudoscalar pole contribution to hadronic light-by-light scattering in the muon $(g-2)_\mu$. Adapted from Ref.~\cite{Gerardin:2016cqj}. \label{fig:HLbL pole dominance}}
\end{figure}

The assumption of hadronic light-by-light scattering being
dominated by single pseudoscalar meson exchange
can be used to calculate the correspondingly leading 
pseudoscalar pole contributions 
$a_\mu^{P\textrm{-pole}}$
to the muon anomly at next-to-leading order (NLO), cf.~\fig\ref{fig:HLbL pole dominance}.
The pole contributions are given by a three-dimensional
integral derived in Ref.~\cite{Knecht:2001qf}.
It
takes the form 
\begin{align}
a_\mu^{P\textrm{-pole}} = \left( \frac{\alpha}{\pi} \right)^3 \int_0^\infty & dQ_1\int_0^\infty dQ_2 \int_{-1}^{+1} d\tilde\tau \nonumber \\ &\bigg[ w_1(Q_1, Q_2, \tilde\tau)  \FPtogg(-Q_1^2, -(Q_1+Q_2)^2) \FPtogg(-Q_2^2, 0) \nonumber \\
	&+w_2(Q_1, Q_2, \tilde\tau)  \FPtogg (-Q_1^2, -Q_2^2) \FPtogg (-(Q_1+Q_2)^2, 0) \bigg] \, ,
   \label{eq:3drepresentation}
\end{align}
where the nonperturbative information is encapsulated in the
transition form factors $\FPtogg$ of the pseudoscalar mesons
$P=\pi^0, \eta, \eta'$ to two virtual photons. The evaluation of the
integrands in \eq(\ref{eq:3drepresentation}) requires the knowledge of
the transition form factors (TFFs) at space-like momenta, both in the
single and double virtual case. It turns out that these TFFs can
indeed be obtained from a QCD calculation on a Euclidean lattice. The
relevant kinematic region is determined by the positive weight
functions $w_1$ and $w_2$ which depend on the absolute values of the
photon momenta, the
kinematic variable $\tilde \tau = \cos \theta \in [-1,+1]$, with
$\theta$ being the angle between the photon momenta, and the mass of
the pseudoscalar
meson $P$. In these proceedings
we focus on the pion-pole contribution for which first lattice results
were obtained in \cite{Gerardin:2016cqj,Gerardin:2019vio}.



\section{The transition form factors on the lattice}
In the continuum Minkowski space the TFFs are defined via the matrix
element of two electromagnetic currents $j_\mu$ and $j_\nu$
and the pseudoscalar state $P$ with four-momentum $p$,
\begin{align*}
M_{\mu \nu}(p, q_1) &= i \int d^4x \, e^{i q_1 x} \left<0\left|T\{j_\mu(x) j_\nu(0) \}\right|P(p)\right> \\
	&= \varepsilon_{\mu \nu \alpha \beta} q_1^\alpha q_2^\beta
   \FPtogg(q_1^2, q_2^2)\, .
\end{align*}
For virtualities below the threshold for hadron production, the
transition form factors can be analytically continued to Euclidean
space, cf.~Ref.~\cite{Gerardin:2016cqj}, and are therefore accessible
on the lattice.
The Euclidean matrix element $M_{\mu \nu}^E(p, q_1)$
can be calculated via an integral over the temporal separation $\tau = t_i - t_f$
of the two currents,
%
\begin{equation}
M_{\mu \nu}^E = \int_{-\infty}^{\infty} d\tau \, e^{\omega_1 \tau} \tilde{A}_{\mu \nu}(\tau), \quad 
i^{n_0} M_{\mu \nu}^E(p, q_1) = M_{\mu \nu}(p, q_1).
\label{eq:Atilde integration}
\end{equation}
Here, $n_0$ denotes the number of temporal indices in $M_{\mu \nu}$, $q_1$ and $q_2$ are the photon virtualities, $p = q_1 + q_2$ is the on-shell pseudoscalar momentum,
$\omega_1$ is a real-valued free parameter with $q_1 = (\omega_1, \vec{q}_1)$, and
\begin{equation*}
\tilde{A}_{\mu \nu}(\tau) = \left<0\left|T\{j_\mu(\vec{q}_1, \tau)
  j_\nu(\vec{p}-\vec{q}_1, 0) \}\right|P(p)\right> \, .
\end{equation*}
On the lattice this function is recovered from the three-point
function
\begin{equation}
C_{\mu \nu} (\tau, t_P) = a^6 \sum_{\vec{x}, \vec{z}} \langle
j_\mu(\vec{x}, t_i) j_\nu(\vec{0}, t_f) P^\dagger(\vec{z}, t_0) e^{i
  \vec{p} \vec{z}}\rangle  e^{-i \vec{x} \vec{q}_1} \equiv \langle j_\mu j_\nu P^\dagger \rangle\, , \label{eq:amplitude}
\end{equation}
via
\begin{equation}
\tilde{A}_{\mu \nu}(\tau) = \frac{2 E_P}{Z_P}\lim_{t_P\to\infty}
e^{E_P (t_f - t_0)} C_{\mu \nu} (\tau, t_P) \,,
\label{eq:Atilde}
\end{equation}
where $t_P = \min(t_f - t_0, t_i - t_0)$ is the minimal temporal
separation between the pseudoscalar and the two vector currents. The pseudoscalar meson energy $E_P$ and the factors $Z_P$ are determined
through appropriate pseudoscalar two-point functions. Before 
integrating over $\tau$, one can contract the Lorentz structure of the
matrix elements. The function $\tilde A_{\mu \nu}$ with one or more temporal indices
vanishes for the pseudocalar at rest, and the spatial components can be written as $\tilde A(\tau) = i
m_P^{-1} \varepsilon_{ijk} \frac{\vec{q}_1^i}{\vec{q}_1^2}
\tilde A_{jk}(\tau)$, and analogously for $C(\tau)$. 

\begin{figure}
\centering
\includegraphics[width=0.3\textwidth]{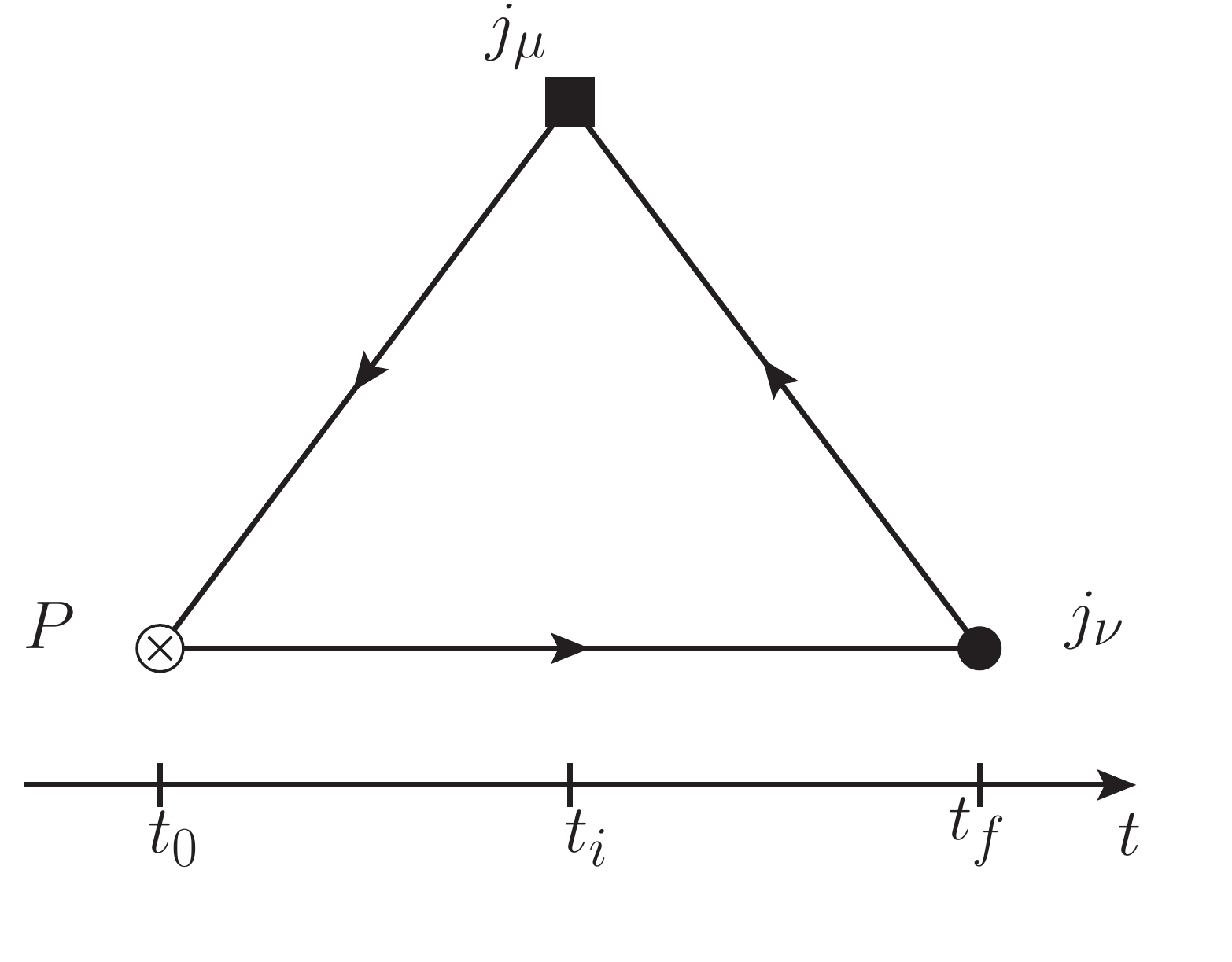}
\includegraphics[width=0.65\textwidth]{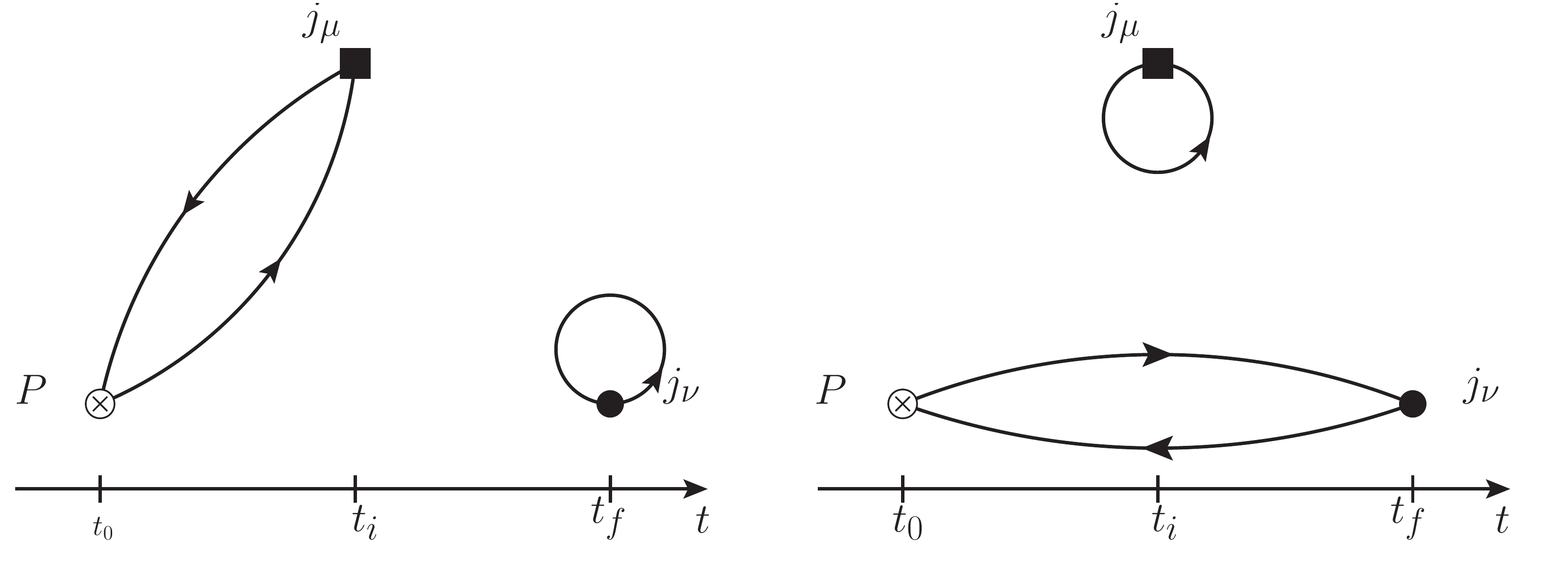}
\includegraphics[width=0.3\textwidth]{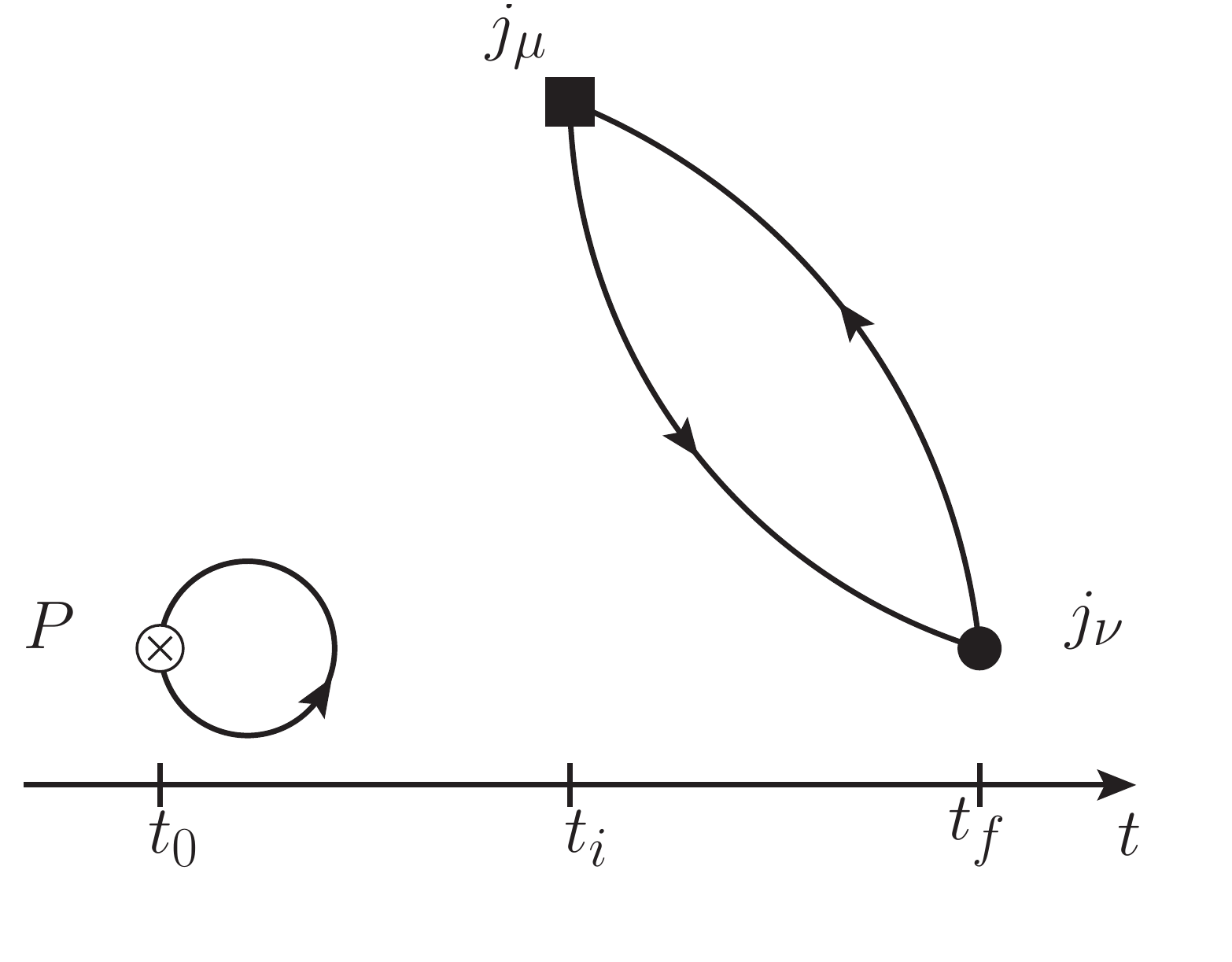}
\includegraphics[width=0.3\textwidth]{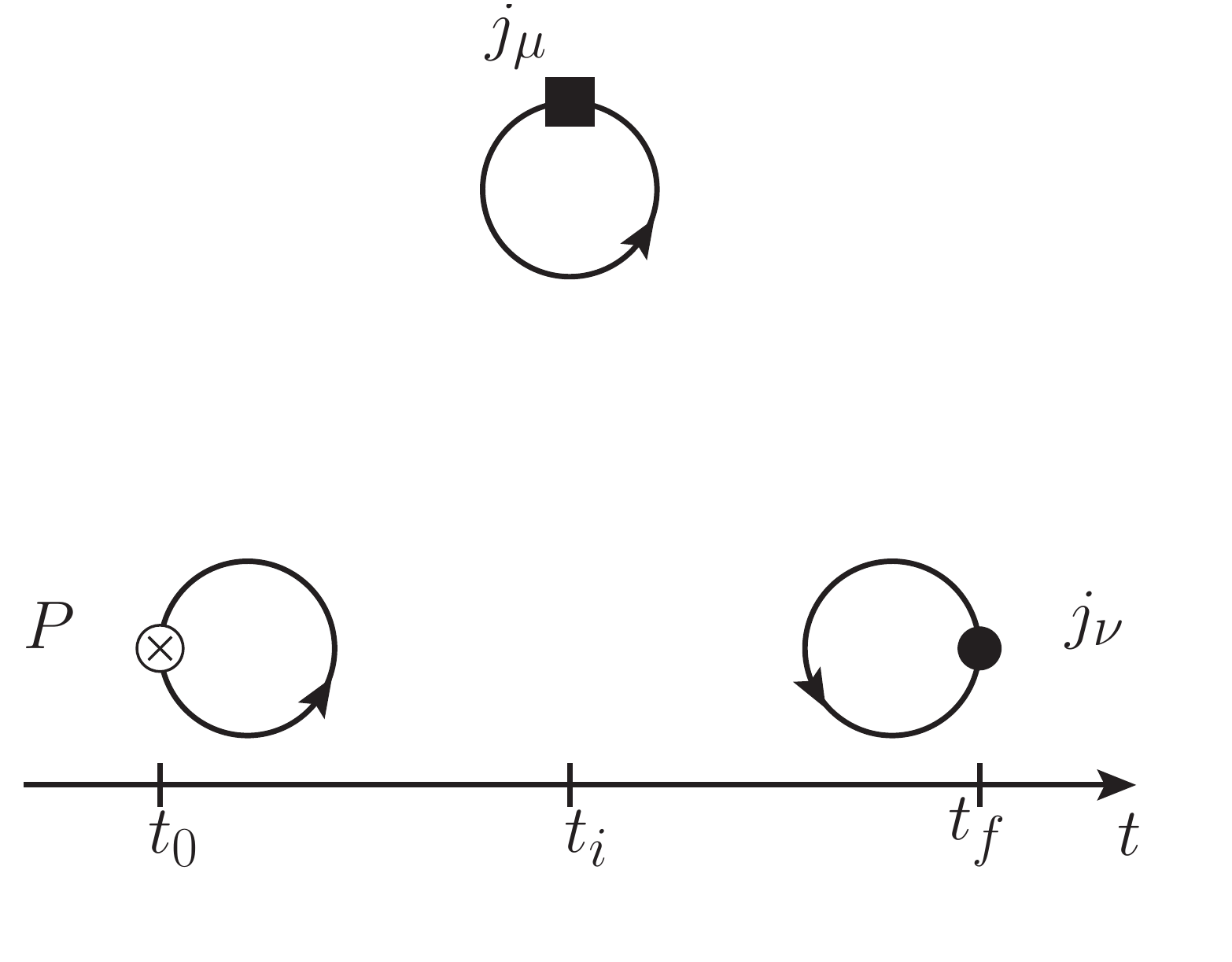}
\caption{Contributions to the three-point function $C_{\mu \nu}$:
  Connected (top left), vector current disconnected (top middle and
  right), pseudoscalar disconnected (bottom left) and fully
  disconnected (bottom right). \label{fig:3pt-diagrams}}
\end{figure}

The amplitude $C_{\mu \nu}$ contains connected, vector current disconnected, pseudoscalar
disconnected, and fully disconnected 
diagrams as illustrated in \fig\ref{fig:3pt-diagrams}.
For Wilson fermions the pseudoscalar disconnected diagrams on
the second line are zero for
$P=\pi_0$ by the exact cancellation between the up and down quark
loops. For $P=\eta$ and $\eta'$ this is not the case and these
disconnected diagrams must be included. This is so also for
$P=\pi_0$ in the twisted mass Wilson fermion discretization, where the
diagrams on the second line are nonzero due to the broken isospin
symmetry.
Since this isospin breaking is a lattice
artefact, we consider an isospin rotation $\pi^0 \rightarrow -i \cdot (\pi^+ + \pi^-)$
with a corresponding transformation of the isospin decomposed light
quark electromagnetic currents $j^{0,0}_\mu \rightarrow j^{0,0}_\mu$
and $j^{1,0}_\mu \rightarrow i \cdot (j^{1,+}_\mu - j^{1,-}_\mu)$,
which allows us to relate the neutral and charged pion form
factors. The difference between the two at finite lattice spacing is a
lattice artefact of order $O(a^2)$.

\begin{wrapfigure}[]{r}{0.5\textwidth}
  \vspace*{-0.7cm}
  \centering
  \includegraphics[width=0.45\textwidth]{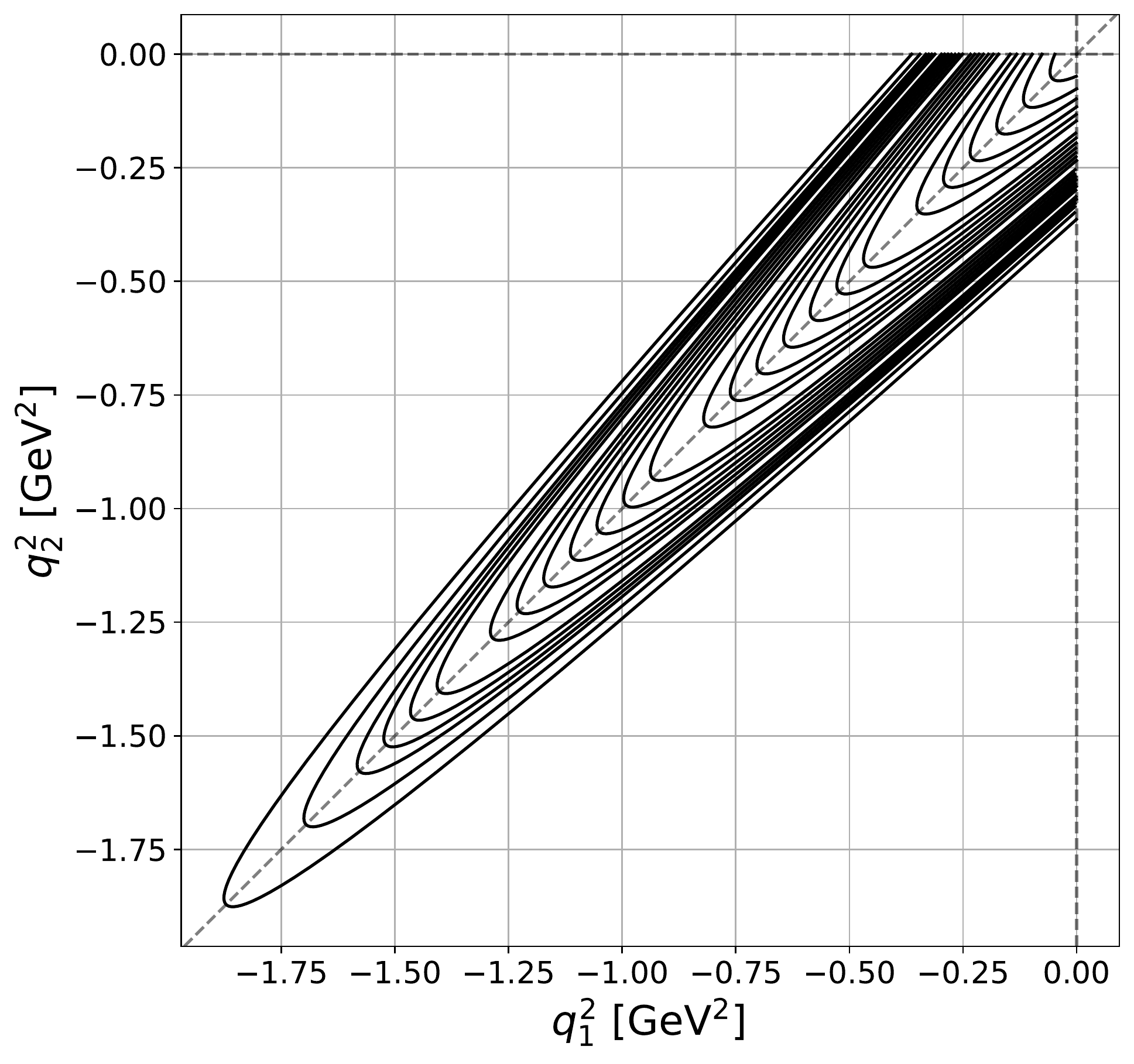}
  \caption{Range of photon virtualities spanned in our calculation on the ensemble cB072.64. \label{fig:yield_plot_cB_phys}}
\end{wrapfigure} 
 A further simplification is achieved by restricting the
considerations to the kinematic situation where the
pseudoscalar is at rest, i.e., $\vec{p} = \vec{0}$. Then,
the expressions for the photon virtualities simplify to
\begin{equation}
q_1^2 = \omega_1^2 - \vec{q}_1^2\,, \quad  
        q_2^2 = (m_P - \omega_1)^2 - \vec{q}_1^2  \, .        \label{eq:orbits}
\end{equation}
%
%
%
%
\noindent
As a consequence, for each choice of spatial momentum $\vec q_1$ one
obtains a continuous set of combinations of $q_1$ and $q_2$ which form
an orbit in the $(q_1^2,q_2^2)$-plane as illustrated in \fig\ref{fig:yield_plot_cB_phys}
for $m_P$ set to the physical pion mass.
There we show the orbits for all the
momenta calculated on the ensemble cB072.64.
From \eqs(\ref{eq:orbits}) it becomes clear that the shape of the
orbits becomes squeezed along the diagonal as the pseudoscalar mass
$m_P$ is lowered. This feature makes it particularly challenging to
extract single virtual pion transition form factors
$\Fpitogg(q_1^2,0) = \Fpitogg(0,q_2^2)$ at large momenta $q_i^2$ on
physical point ensembles if one uses only pions at rest. However, the problem
can be circumvented by using moving frames, 
cf.~\cite{Gerardin:2019vio}. For $P=\eta$ and $\eta'$ the problem is
less eminent due to the larger values of the meson masses $m_P$.

\section{First results at the physical point}
\begin{figure}[htbp]%
\centering
\includegraphics[page = 1, width = 0.49\textwidth]{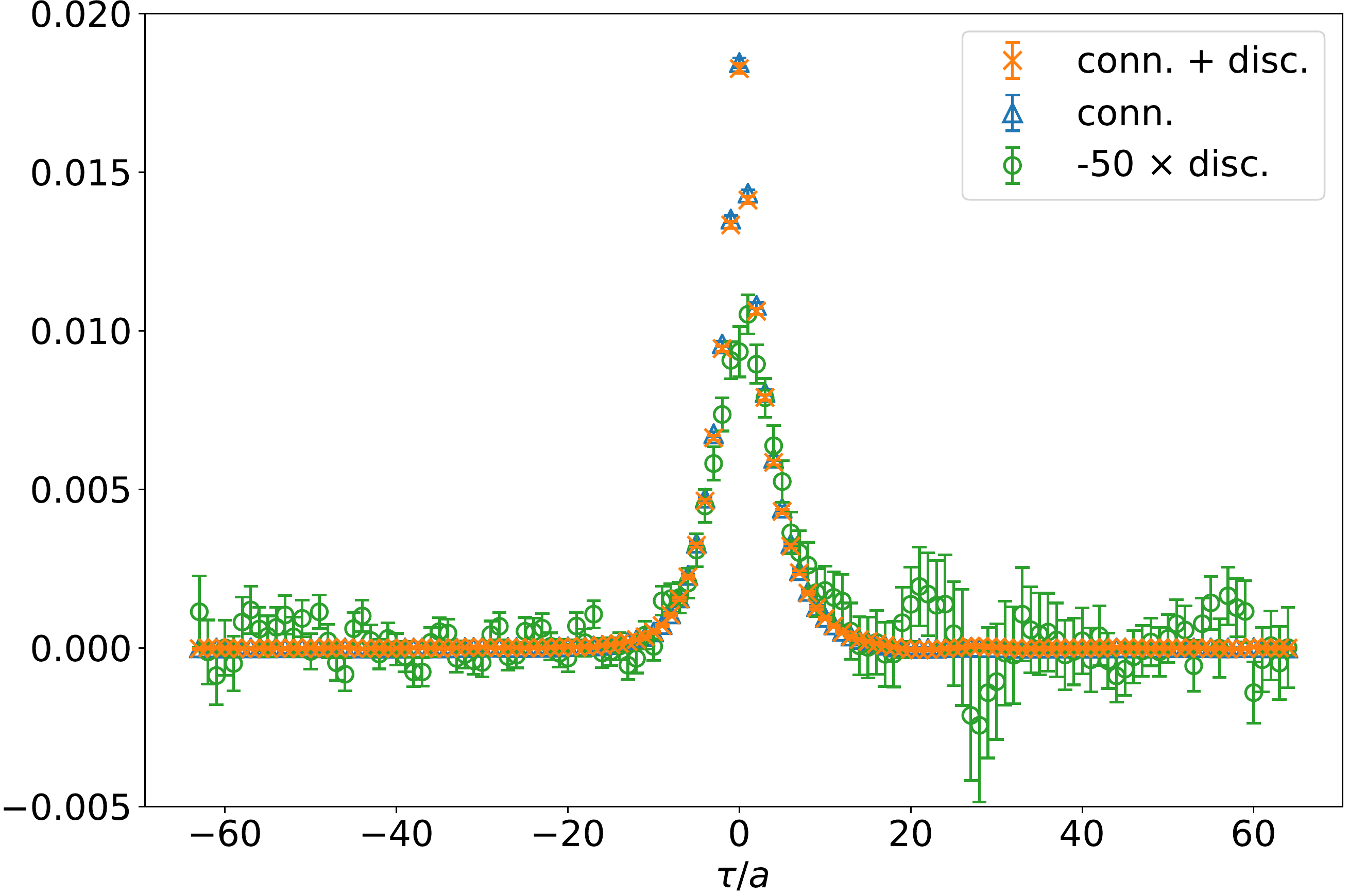}\hfill
\includegraphics[page = 2, width = 0.49\textwidth]{Figures/plots_divided_wraparound_lm+hp_vdisc_comp_cB211}
\caption{Amplitude $\tilde A(\tau)$ for momentum orbit $|\vec{q}^2| =
  10\,(2\pi/L)^2$ (left) and $|\vec{q}^2| = 29\,(2\pi/L)^2$ on
  cB072.64. Shown in orange is the full contribution to $\tilde A(\tau)$, in blue the
  connected contribution and in green the vector current
  disconnected contribution multiplied by -50.\label{fig:plot_Atilde}
}
\end{figure}

After this theoretical discussion we are now in the position to
present first results for the transition form factor $\Fpitogg$ of the
pion obtained for the ensembles cB072.64 and cC060.80 at the physical point.
First, we illustrate the quality of our data with sample results 
for the amplitude $\tilde A(\tau)$ defined in
\eq(\ref{eq:Atilde}).
In \fig\ref{fig:plot_Atilde} we show 
the full amplitude and separately the fully
connected and the vector current disconnected contributions 
for two of the momentum orbits on the ensemble
cB072.64. 
The vector current disconnected amplitude is multiplied by a factor $-50$ in order to facilitate comparison
with the connected contribution and the full amplitude.
%
The examples illustrate that the
disconnected contribution is very small, but significant. More
generally, we find that in the peak region it is
suppressed w.r.t.~the connected contribution by a factor between 50
and 200 depending on the orbit. We also conclude from our data that the
statistical error on the disconnected contribution is sufficiently well under control on the physical
point ensembles.

\begin{figure}[htbp]
  \includegraphics[page=2, width = 0.49\textwidth]{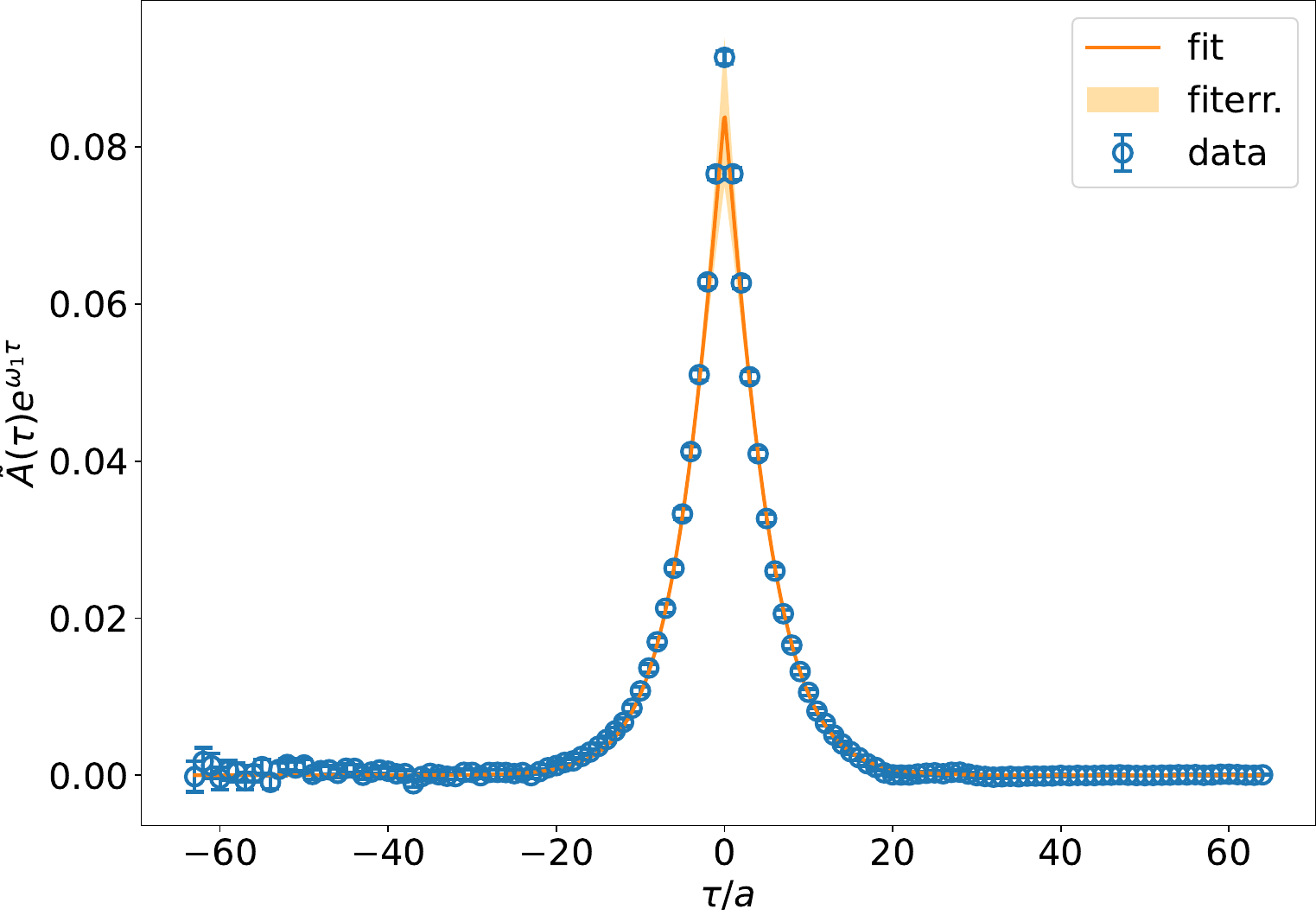} \hfill
  \includegraphics[page=2, width = 0.49\textwidth]{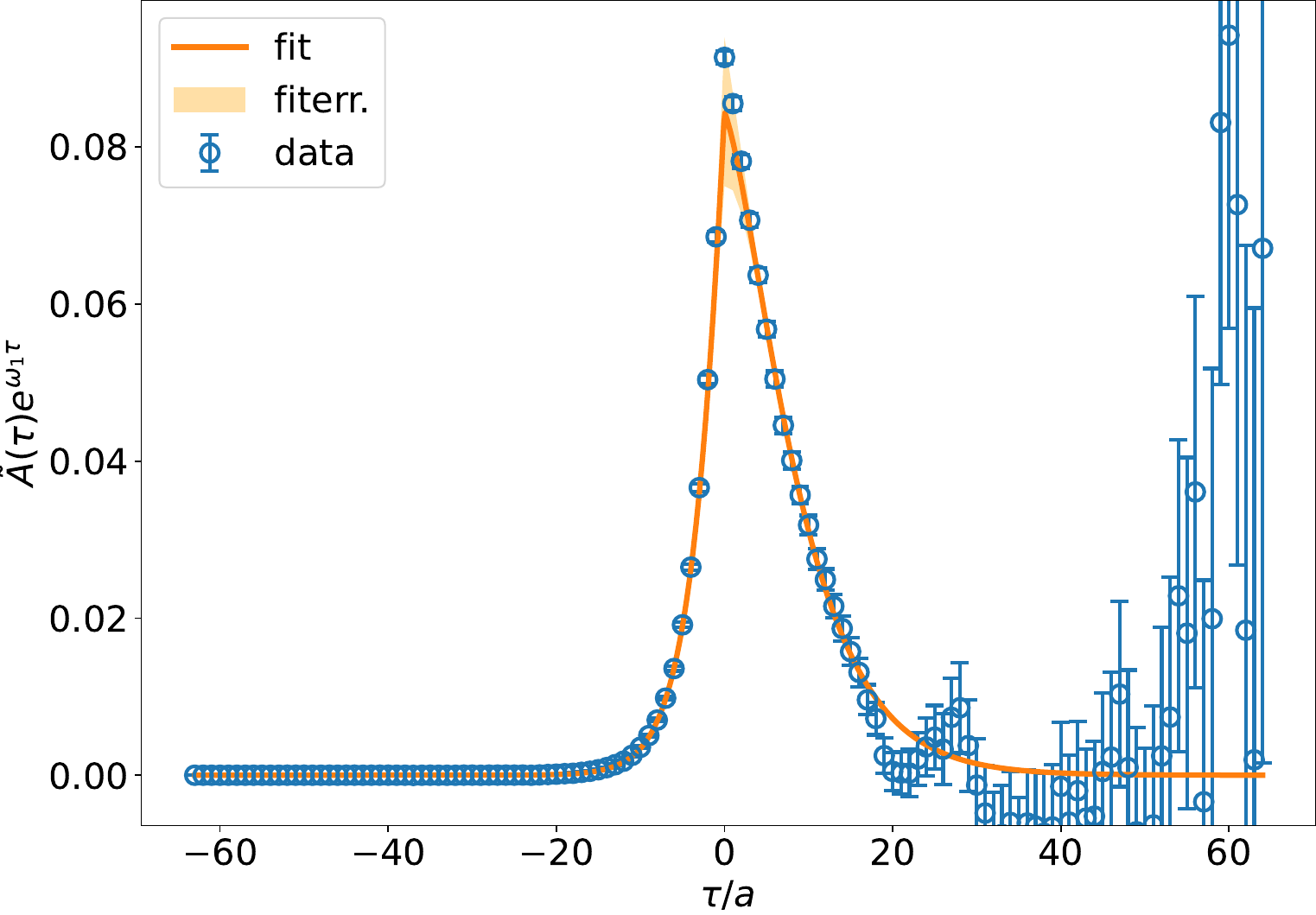}
  \caption{Integrand $\tilde A(\tau) e^{\omega_1 \tau}$ on cB072.64
    with LMD model fits for momentum orbit $|\vec{q}^2| =
    2\,(2\pi/L)^2$. Diagonal kinematics with $a \omega_1 = a m_\pi/2
    \approx 0.0284$ (left), single virtual kinematics with $a \omega_1
    = a |\vec{q_1}| \approx 0.1388$ (right). \label{fig:global_LMD_fit} }
\end{figure}

To obtain the form factor we need to integrate $\tilde
A(\tau)$ weighted by the factor $\exp(\omega_1 \tau)$ over the whole temporal axis, cf.~\eq\ref{eq:Atilde integration}. In order to control the
statistical error in the exponentially enhanced tail and to be able to
integrate up to $\tau \rightarrow \infty$, we proceed as
follows. First, we fit the lattice data by a model function $\tilde
A^\textrm{(fit)}(\tau)$ in a range
$\tau_\textrm{min} \leq |\tau| \leq
\tau_\textrm{max}$, and then we
replace the lattice data $\tilde
A^\textrm{(latt.)}(\tau)$ by the data
from the fit for $\tau > \tau_\textrm{cut}$,
\begin{equation}
  \Fpitogg(q_1^2,q_2^2) = \int_{-\infty}^{\tau_\textrm{cut}} d\tau \, \tilde A^\textrm{(latt.)}(\tau) e^{\omega_1 \tau} + \int_{\tau_\textrm{cut}}^{\infty} d\tau \, \tilde A^\textrm{(fit)}(\tau) e^{\omega_1 \tau}. \label{eq:TFF_fit}
\end{equation}
Following Ref.~\cite{Gerardin:2016cqj} we use both a vector meson
dominance (VMD) model and
the lowest meson dominance (LMD) model to estimate the model
dependence. We perform global fully correlated fits, i.e., we simultaneously fit
all momentum orbits in the range $\tau_\textrm{min} \leq |\tau| \leq
\tau_\textrm{max}$ and take into account the correlation between all
fitted data. In \fig\ref{fig:global_LMD_fit} 
we illustrate the procedure by showing the result for the integrand $\tilde A(\tau)
e^{\omega_1 \tau}$ of a typical global
fit to $\tilde A(\tau)$ in the range $9 \leq |\tau/a| \leq 12$ with $\chi^2/\textrm{dof}=1.20$ on the ensemble cB072.64 using the LMD model. The plot on the left shows the resulting
integrand for the diagonal kinematics $q_1^2=q_2^2$,
while the plot on the right shows it for the single
virtual kinematics with $q_1^2=0$.
The transition form
factors obtained from the integration over the lattice data and the fitted
data depend of course on the choice of the model, the fit range and
the value $\tau_\textrm{cut}$. The variations resulting from
these choices are carried through all further analysis steps and are
included in the systematic error estimate of the final result for
$a_\mu$. The typical values of $\tau_\textrm{cut}$ we use in our
analysis result in a data content of well above 98\% for most of the
TFFs. However, for TFFs with (close to) single virtual kinematics, the
data content is sometimes also less for higher momentum orbits.
Here, the data content is defined as the fraction of the TFF
coming from the first term in Eq.~(\ref{eq:TFF_fit}). 

\begin{figure}[htbp]%
\centering
\includegraphics[page=1, width =
1.0\textwidth]{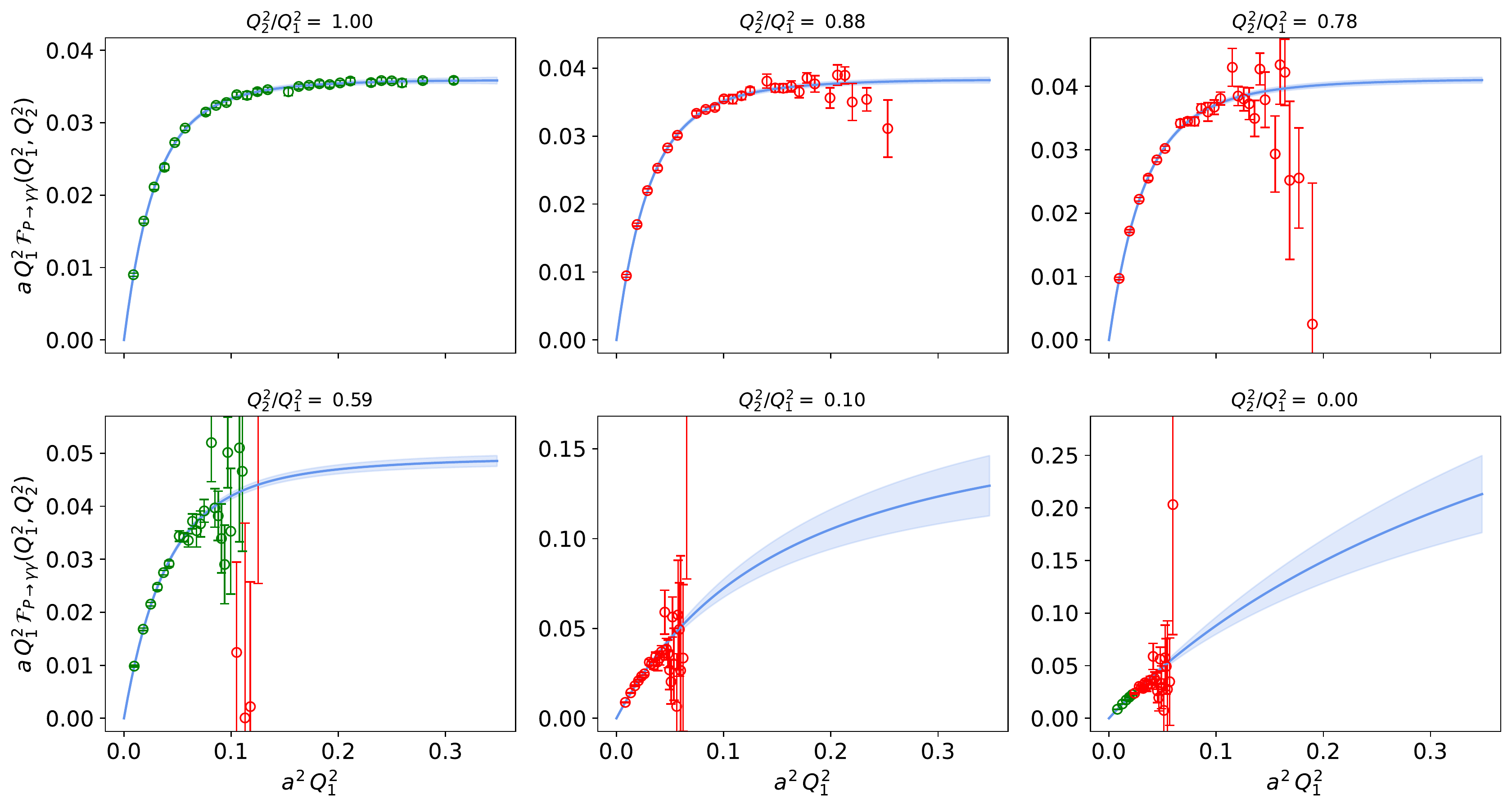}
\caption{Illustration of transition form factors and their
  parameterization using the fitted modified $z$-expansion. Only the data
coloured in green is included in the fit. 
\label{fig:zExpansion}}
\end{figure}

Once the form factors are obtained in the whole kinematic region as
described by the yield plot in \fig\ref{fig:yield_plot_cB_phys}, we
parameterize them using a modified $z$-expansion of the form 
\begin{multline}
P(Q_1^2, Q_2^2) \cdot \mathcal{F}_{\pi \gamma^* \gamma^*} (-Q_1^2, -Q_2^2) = \\
 \sum_{m,n = 0}^N c_{nm} \left(z_1^n - (-1)^{N+n+1} \frac{n}{N+1}
   z_1^{N+1} \right) \left(z_2^m - (-1)^{N+m+1} \frac{m}{N+1}
   z_2^{N+1} \right)
 \label{eq:modified z-expansion}
\end{multline}
where $z_k = z(Q_k^2)$ are modified four-momenta and $P(Q_1^2, Q_2^2)$
is a polynomial, see Ref.~\cite{Gerardin:2019vio} and references therein for
further details. We determine the coefficients $c_{nm}$ by fitting
\eq(\ref{eq:modified z-expansion}) to samples
of $\Fpitogg(-Q_1^2,-Q_2^2)$ in the $(Q_1^2,Q_2^2)$-plane. The sample
points are given by a set of fixed values of $Q_2^2/Q_1^2$ on all
momentum orbits, and we ensure that all included data points pass a
certain 
threshold for the data content.
In \fig\ref{fig:zExpansion} we show the
result of such a (fully correlated) fit with
$\chi^2/\textrm{dof}=0.96$ using $Q_2^2/Q_1^2=1.0, 0.59, 0.0$,
and $N=2$ to the TFFs obtained from a global LMD fit with $\{\tau_\textrm{min}/a,
\tau_\textrm{max}/a\} = \{9, 12 \}$, $\chi^2/\textrm{dof}$ = 1.20,
$\tau_\textrm{cut}/a$ = 20 and a threshold of 90\% on the ensemble cB072.64. 
As a crosscheck for the quality of the
fit we also show the data for three other ratios $Q_2^2/Q_1^2=0.88,$ 0.78, and
0.10 not included in the fit together with the fitted modified $z$-expansion. The variations resulting from
varying the sampling of $\Fpitogg(-Q_1^2,-Q_2^2)$ in the momentum plane are also
included in the systematic error estimate of the final result for
$a_\mu$.

\begin{figure}[htbp]%
\centering
\includegraphics[width=0.49\textwidth]{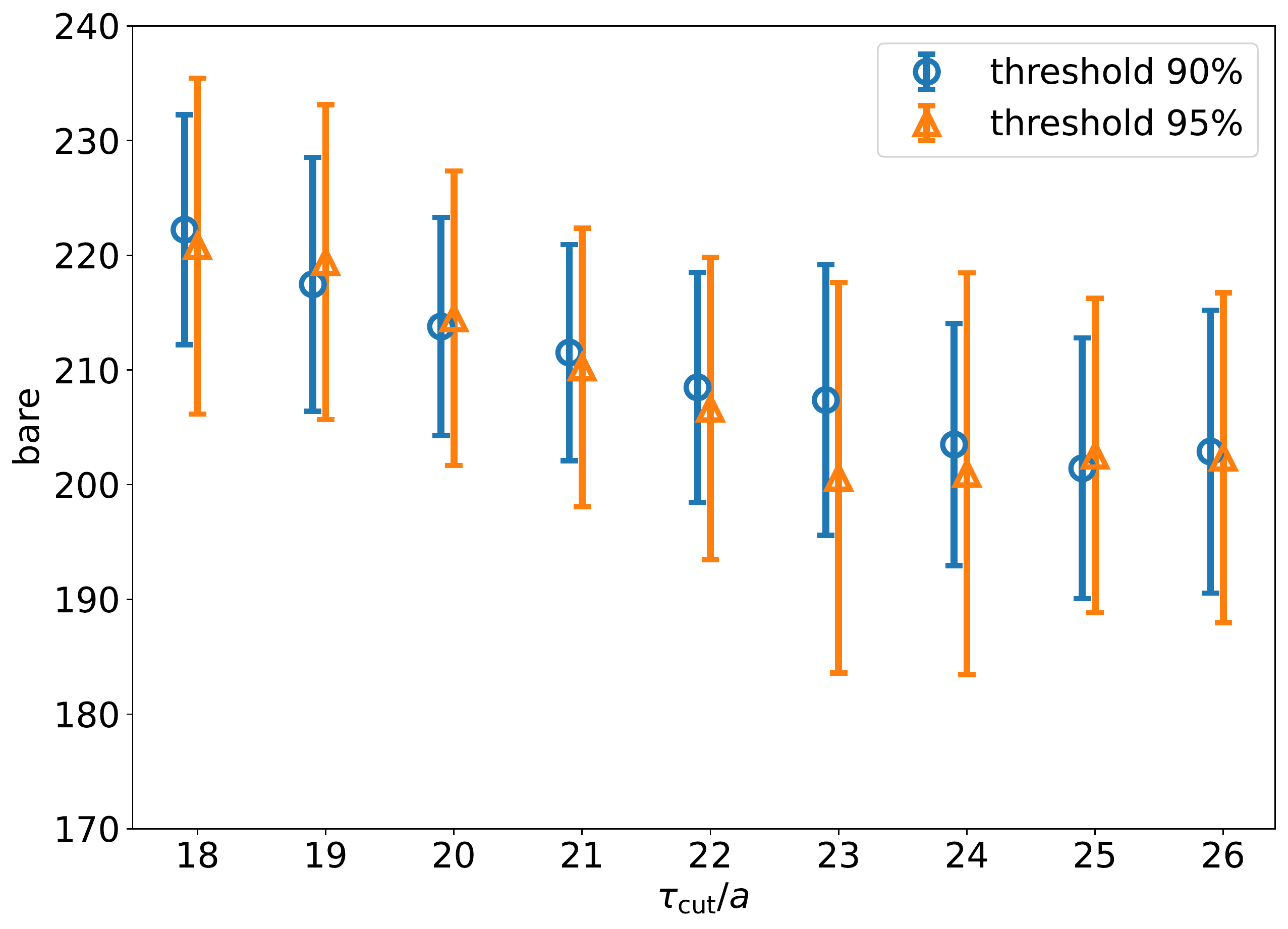}\hfill
\includegraphics[width=0.49\textwidth]{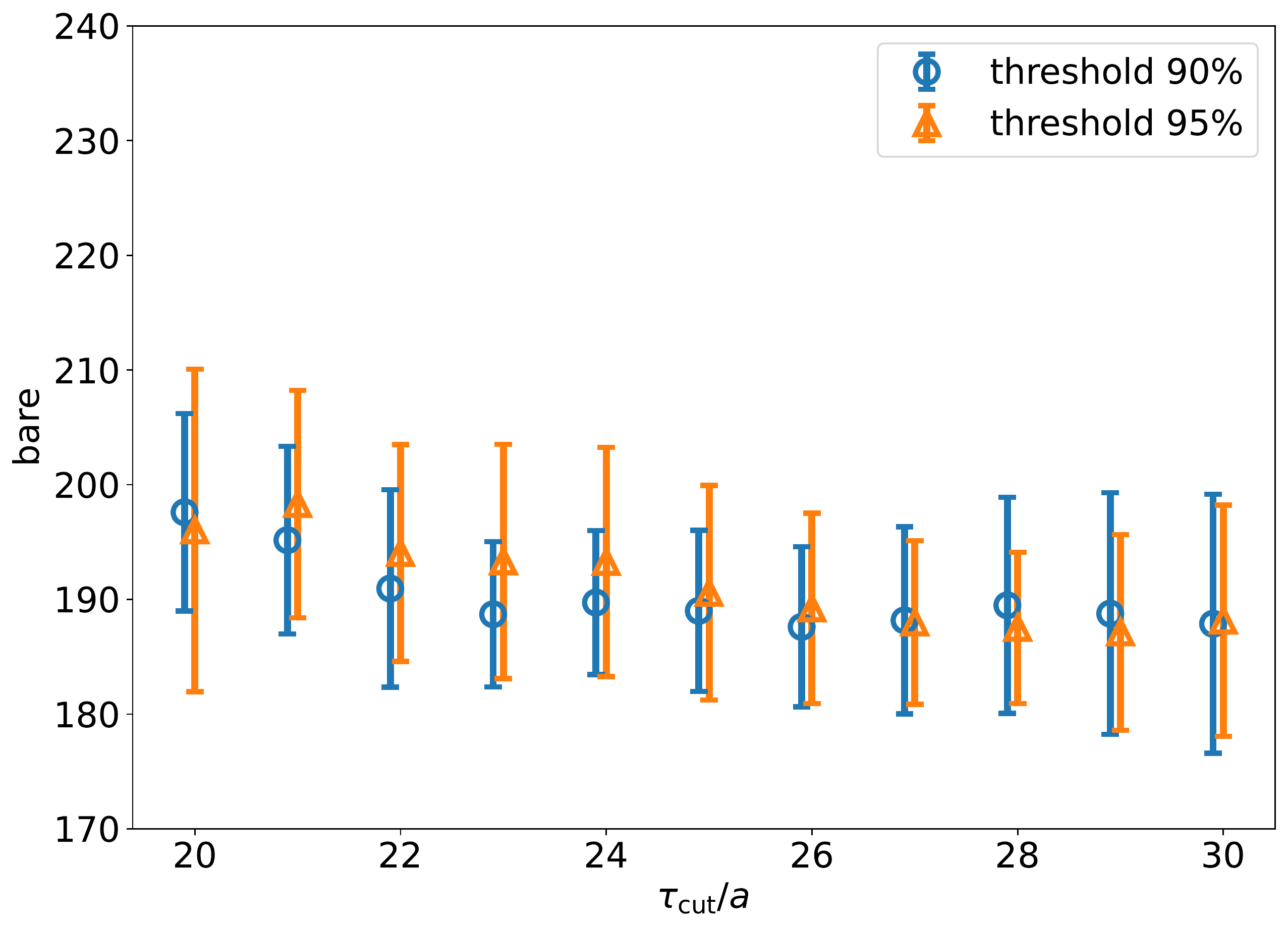}
\caption{AIC averaged data for a range of $\tau_\textrm{cut}/a$ for the ensembles cB072.64 (left) and cC060.80 (right). \label{fig:tc-dependence}}
\end{figure}

Finally, having the parameterization of the TFFs at hand, we can use
it in the three-dimensional integral representation in
\eq(\ref{eq:3drepresentation}) and calculate the bare pion-pole
contribution $a_\mu^{\pi\textrm{-pole, bare}}$ to the anomalous
magnetic moment. In \fig\ref{fig:tc-dependence} we show the results
for $a_\mu^{\pi\textrm{-pole, bare}}$ on the two ensembles at the
physical point as a function of $\tau_\textrm{cut}/a$. Each data point
is a weighted average of $O(100)$ results from
different fits for $\tilde A$ using VMD or LMD with different fit
ranges and different fits using the modified $z$-expansion on different
samplings in the momentum plane. The weighted average is obtained
using weights inspired by the Akaike information criterion (AIC). The error
therefore includes the variation w.r.t.~the fitting of $\tilde A$ and
the sampling of $\Fpitogg$ in the $(Q_1^2,Q_2^2)$-plane.
The variation of the final result with $\tau_\textrm{cut}$
indicates a residual dependence on the specific procedure of variance
reduction in the large-$\tau$ tail of $\tilde A$. In principle, this dependence is
removed in the limit $\tau_\textrm{cut} \rightarrow \infty$, but if $\tau_\textrm{cut}$
is chosen too large the
$z$-expansion fits become unstable and hence the final
result unreliable. Our results in
Figure \ref{fig:tc-dependence} indicate that choosing $\tau_\textrm{cut} \in [1.8, 2.1]$ fm
seems a safe choice and we
perform a further AIC averaging over this range. This yields the bare results shown in Table \ref{tab:bare_results} for the two physical point ensembles, with total errors in the 5\%-8\% range.
\begin{table}[t]
\centering
\begin{tabular}{| l | l | l |}
\hline
$a_\mu^{\pi-\text{pole, bare}} \cdot 10^{-11}$ & threshold 90\% & threshold 95\% \\ \hline
cB072.64 & 208.9(10.1)(7.8)[12.8] & 204.5(14.2)(6.3)[15.6] \\
cC060.80 & 188.9(9.9)(2.7)[10.2] & 187.9(9.0)(1.9)[9.2] \\
\hline
\end{tabular}
\caption{Bare results using the AIC procedure on the two physical point ensembles. The first error is the statistical error, the second the systematic error and the third the total error. \label{tab:bare_results}}
\end{table}
Since we use local iso-vector and iso-vector axial current operators
in our amplitude $C_{\mu \nu}$, instead of conserved (point-split)
current operators, we need to renormalize the bare results by the corresponding
renormalization constants. Preliminary values are available for our setup from a calculation within ETMC.

\section{Conclusion and outlook}
After applying the renormalization factors and performing a rough estimate
of the continuum limit, we obtain a preliminary value 
$a_\mu^{\pi-\text{pole}} = 53.7(2.6)(3.1)[4.0] \cdot 10^{-11}$. This
can be compared to the recent lattice result $a_\mu^{\pi-\text{pole}}
= 59.7(3.6) \cdot 10^{-11}$ from Ref.~\cite{Gerardin:2019vio} and the
dispersive result $a_\mu^{\pi-\text{pole}} = 63.0^{+2.7}_{-2.1} \cdot
10^{-11}$ from Refs.~\cite{Aoyama:2020ynm,Hoferichter:2018dmo,Hoferichter:2018kwz}, and we find agreement
within 1 to 2 standard deviations. Finalizing the analysis might result in a slightly
different central value, however, we expect that the relative total
error will stay below the 10\% level. We plan to analyze a third
physical point ensemble at a finer lattice spacing which will result in
a more robust continuum limit extrapolation. We also plan to
calculate the form factors for the pion in a moving frame and to perform the analysis of the $\eta$- and $\eta'$-pole contributions, and to include ensembles with larger pion masses.


\bibliographystyle{JHEP}
\bibliography{p2gg}



\end{document}